\documentclass[12pt]{article}
\input epsf
\newcommand{\text}{\rm}
\begin{document}

\title{\textbf{Long range interactions and the $3D$ asymptotic fermion spectrum}}

\author{Daniel G. Barci\thanks{Regular Associate of the Abdus Salam International Centre for Theoretical Physics, ICTP, Trieste, Italy.} \\ Departamento de F\'\i sica Te\'orica,\\
Universidade do Estado do Rio de Janeiro,\\Rua S\~ao Francisco Xavier 524, 20550-013,
Rio de Janeiro, RJ, Brazil. \\ \\
Luis E. Oxman\thanks{email: oxman@if.uff.br} \\ Instituto de F\'{\i}sica, Universidade Federal Fluminense,\\ Av. Litor\^anea S/N, Boa Viagem,\\
Niter\'oi, RJ 24210-340, Brazil.\\ }

\maketitle

\begin{abstract}
In this article, we study the stability of the space of asymptotic fermion states in $(2+1)D$, when long range interparticle interactions are present. This is done in the framework of bosonization, where the fermion propagator can be represented in terms of a vortex correlator. In particular, we discuss possible instabilities in the large distance behavior of the induced action for the vortex worldline. \\
\end{abstract}





\renewcommand{\theequation}{\thesection.\arabic{equation}}

\section{Introduction}

In two dimensional spacetime, bosonization enables the identification of fer\-mions with finite energy solitons of a dual sine-Gordon model \cite{b2}. These solitons are topological objects that interpolate between two different vacua $\phi(x\to-\infty)=0$ and $\phi(x\to+\infty)=2\pi$; for the antisoliton the behavior is similar, just interchanging the vacua.
Many important consequences can be easily derived in the dual scenario. In particular, when a fermion interaction coming from the
minimal coupling with a Maxwell term is introduced (Schwinger model) \cite{schwinger}, a mass term $\phi^2$ is induced in the dual theory. As a consequence, a configuration with nonzero topological charge (it tends to a nonzero value either at $+\infty$ or $-\infty$) cannot have finite energy. In other words, a single soliton, or in fermionic language the single fermion, disappear from the space of asymptotic states. This can be understood as due to the confining nature of the Maxwell interaction in two dimensions, which implies a linear potential between charges.

A natural question that arises is whether a possible destabilization of the fermionic spectrum in higher dimensional systems
can be studied in a similar manner. In reference \cite{ox-so}, fermions in three dimensions have been associated with vortex configurations
of the dual bosonized gauge theory. Then, the above mentioned program can be rephrased in terms of possible instabilities of the vortex modes as a
consequence of interactions.

As we will see, in the three dimensional case, interactions do not render in general the vortex energy infrared divergent, and it is harder to display scenarios where the destabilization of the vortices is observed. This situation is expected as we know that while in two dimensions the fermion spectrum is easily destabilized by interactions, in three or more dimensions, the quasiparticle picture is quite robust.

Then, we will search for possible destabilizations associated with the localization properties of the dual model. This aim can be understood in the following manner.  The association of vortices in the bosonized theory and fermion states has been accomplished by using a path integral definition of the vortex propagator, by creating and annihilating the vortex at the position of an instanton
anti-instanton pair in euclidean spacetime \cite{ox-so}. These euclidean objects can be seen as a monopole anti-monopole pair joined by a Dirac string which becomes observable. When the theory presents a well defined localization length, as in the case of free fermions, the Dirac string can be seen as the vortex worldline and at large distances the induced action is dominated by the length of the string plus Polyakov's action for spin one half particles. This leads to a pole in the vortex propagator, associated with the single fermion state. Therefore, a possible breakdown of the above mentioned large distance behavior, induced by the fermion interactions, could eliminate the single fermion from the asymptotic space of states. We will show that, although the part of the induced action proportional to the string length is robust, there is a set of interactions that could destabilize the spin part.

The paper is organized as follows: in section \S \ref{model} we describe a general model of fermions coupled to $U(1)$ gauge fields and discuss the dual theory in  $(1+1)D$ and $(2+1)D$, emphasizing the role of topological excitations.
In section \S\ref{analisis} we analyze possible destabilizations of the fermion propagator due to interactions. Finally, in section
\S\ref{discusion} we discuss our results.

\section{Fermions coupled to $U(1)$ gauge fields
\label{model}}

Let us consider massive fermions coupled with a gauge field,
\begin{equation}
K_{F}[\psi ]+\int d^\nu x\, [{\mathcal A}\, j^{F}+\frac{1}{4}{\mathcal F}_{\mu \nu} \hat{{\mathcal P}}{\mathcal F}_{\mu \nu}]\;,
\label{modelo}
\end{equation}
where $K_{F}$ is the action for free fermions , ${\mathcal F}_{\mu \nu}$ is the field strength tensor for a gauge field,
and we have included a possible nonlocal operator $\hat{{\cal P}}$, associated with a kernel ${\cal P}(x-y)$,
\[
\left[ \hat{{\mathcal P}} {\mathcal F}\right] (x)=\int d^{\nu}y\, {\mathcal P}(x-y) {\mathcal F}(y)\;.
\]
We also define the Fourier transform,
\begin{equation}
\tilde{{\mathcal P}}(k)=\int d^{\nu}x\, e^{ikx} {\mathcal P}(x)\;.  \label{f}
\end{equation}
The usual Maxwell term is recovered by taking $\hat{{\mathcal P}}\equiv I$, that is, ${\mathcal P}(x-y)=\delta^{(\nu)}(x-y)$.

The integration over the gauge fields yields
for the total fermion action,
\begin{equation}
K_{F}[\psi ]+I[j^{F}]
\makebox[.5in]{,}
I[j^{F}]=\frac{1}{2}\int d^\nu x\, j^{F}\frac{1}{(-\partial^2)\hat{{\mathcal P}}}\; j^{F}\;,
\label{KF}
\end{equation}
where $\nu$ is the spacetime dimension.

\subsection{$(1+1)D$ massive fermions \label{1d}}
In the simple case of massive fermions in $(1+1)$D, the model in eq.(\ref{KF}) can be bosonized according to \cite{b2},
\begin{equation}
K_{F}[\psi ]+I[j^{F}] \leftrightarrow K_B[\phi]+ I[\epsilon \partial \phi]\;,
\label{bos1d}
\end{equation}
\begin{equation}
K_B[\phi]=\int d^2x\, \left[\frac{1}{2}\partial_\mu \phi \partial^\mu \phi -\mu \cos \beta \phi\right]
\makebox[.5in]{,}
I[\epsilon \partial \phi]=\frac{1}{2}\int d^2x\, \phi \frac{1}{\hat{{\cal P}}} \phi\;,
\label{mass}
\end{equation}
where we have used the bosonization rule for the fermion current $j^{F}=\epsilon \partial \phi$, or in components,
$j^{F}_\mu=\epsilon_{\mu \nu} \partial_\nu \phi$. When the interaction induced by the gauge fields is absent, the model contains the well known solitons representing the fermions in the bosonized language. These solitons are well localized finite energy solutions with nonzero topological charge. When $\hat{{\mathcal P}}\equiv I$, the mass term in eq. (\ref{mass}) implies that configurations with nonzero topological charge cannot have finite energy, as they are different from zero either at spatial $+\infty$ or $-\infty$. In other words, the Maxwell interaction in $(1+1)$D eliminates the single soliton modes. This can be understood as in this case the fermions experience a linear interaction.

An interesting question arises if we consider a general Lorentz invariant kernel $\hat{{\mathcal P}}={\mathcal P}(\partial^2)$ in eq. (\ref{mass}). Is there any  situation in which the generally nonlocal $\phi^2$ term could allow for finite energy solutions in the infrared regime? For this aim, we can consider a static field,
\begin{equation}
\phi(\mathbf{x})=\int_{-\infty}^{+\infty} d\mathbf{k}\, \tilde{\phi}(\mathbf{k})\exp{(-i\mathbf{k}\mathbf{x})}\;,
\label{fourier}
\end{equation}
with nonzero topological charge, thus satisfying,
\begin{equation}
\int_{-\infty}^{+\infty} d\mathbf{x}\, \partial_\mathbf{x} \phi (\mathbf{x}) \neq 0\;,
\end{equation}
this means $\tilde{\phi}(\mathbf{k})\sim 1/\mathbf{k}$, for $\mathbf{k}\to 0$. On the other hand, using the Fourier transformed variables, the nonlocal $\phi^2$ term reads,
\begin{equation}
\int d\mathbf{k}\, \frac{1}{\tilde{\mathcal{P}}(\mathbf{k})}|\tilde{\phi}(\mathbf{k})|^2\;.
\end{equation}
Therefore, considering the class of operators having
\begin{equation}
\hat{{\mathcal P}}(k)=\lambda/k^\alpha\;,
\label{alfa}
\end{equation}
the energy of a nonzero topological configuration will be infrared finite whenever $\alpha > 1$. The static potential $V(\mathbf{x})$ associated with this nonlocal operator,
\begin{equation}
\hat{{\mathcal P}}\partial_\mathbf{x}^2  V=\delta (\mathbf{x})\;,
\end{equation}
corresponds to $V(\mathbf{x})\propto |\mathbf{x}|^{1-\alpha}$. Then, the above mentioned energy is infrared finite whenever the nonlocal character of the Maxwell term drives the static potential into a nonconfining one.

\subsection{Bosonization and the vortex propagator in $(2+1)D$ \label{2d}}
Let us consider a similar situation in $(2+1)D$, where it is also well
established that the correlation functions of $U(1)$ fermionic currents
correspond to correlation functions of topological currents in the dual
bosonized theory \cite{b,b11}. This feature has a universal character, generalizing eq. (\ref{bos1d}). In other words, we have the following formula
\cite{b111,b1},
\begin{equation}
K_{F}[\psi ]+I[j^{F}]\leftrightarrow K_{B}[\lambda ]+I[\varepsilon \partial
\lambda ]  \label{univ}
\end{equation}
where, $K_{B}$ is the corresponding bosonized action and the bosonizing field $\lambda$
is a scalar field $\phi $ in $(1+1)D$, and a gauge field $A_{\mu }$ in $(2+1)D$. Accordingly,
$\varepsilon \partial \lambda $ has to be read as $\varepsilon _{\mu \nu }\partial _{\nu }\phi $
or $f_\mu =\varepsilon _{\mu \nu \rho }\partial _{\nu }A_{\rho }$, respectively.

When parity breaking fermions in $(2+1)D$ are considered, and a large mass expansion is
performed, the dominant part of $K_{B}$ reduces to the Maxwell-Chern-Simons (MCS) model \cite{b,b11},
\begin{equation}
K_B[A]\sim \int d^{3}x\left( \frac{1}{2m} f_{\mu }^{2}\;+\;\frac{i}{2\eta}A_{\mu }f^{\mu }\;\right) \;,  \label{mcs-act}
\end{equation}
where $m$ is proportional to the fermion mass and $\eta $ is the
Chern-Simons coefficient in the fermion effective action.
Using the universal mapping (\ref{univ}), the bosonized form of the total fermion action in eq. (\ref{KF}), including
interactions, turns out to be,
\begin{equation}
S[A]=\int d^{3}x\left( \frac{1}{2} f_{\mu }\hat{O} f_{\mu } \;+\;%
\frac{i}{2\eta }A_{\mu }f^{\mu }
\;\right) \;,  \label{sec-act}
\end{equation}
\begin{equation}
\hat{O}=\frac{1}{m}+\frac{1}{(-\partial^2)\hat{{\mathcal P}}}\;.
\label{valoro}
\end{equation}

In a similar manner to the $(1+1)D$ case, where free fermions can be associated with soliton
configurations in the dual massive sine-Gordon model, in the case of parity
breaking matter in $(2+1)D$, they can be associated with vortices of the bosonized dual theory.

Following 't Hooft procedure \cite{th}, the vortex propagator in euclidean spacetime is
obtained by path integrating over configurations where a vortex excitation is
created out of the vacuum at a space-time point $x_{1\mathrm{\ }}$ and after
an intermediate propagation is annihilated at $x_{2}$. Before $x_{1}$ and
after $x_{2}$ the topological charge vanishes, while it is nonvanishing in
between due to the existence of the vortex. Therefore, suitable instanton
anti-instanton singularities have to be introduced at $x_{1}$ and $x_{2}$ in
order to match these inequivalent topological configurations. In the present
three dimensional case these singularities can be seen as a monopole
anti-monopole pair \cite{ht,pi} for the dual field strength $f_{\mu }$, located
at $x_{1}$ and $x_{2}\,$, respectively. This pair is associated with a Dirac string
$\gamma$ running from $x_{1}$ to $x_{2}$, which becomes observable in the MCS model.

Therefore, the two-point vortex correlation function is defined by,
\begin{equation}
\mathcal{G}(x_{1}-x_{2})=\int D\gamma \int DA\;e^{-S[A,J]}=\int D\gamma
\;e^{-\Gamma _{\gamma }\;}\;,  \label{corr}
\end{equation}
where $S[A,J]$ represents the coupling of the string and the MCS model, and $\Gamma _{\gamma }$ is
the induced string action obtained by integrating over all gauge configurations in a fixed string background.
As discussed in ref. \cite{ox-so}, for the coupling,
\begin{equation}
S[A,J]=\int d^{3}x\left( \frac{1}{2m}(f_{\mu }+J_{\mu })\hat{O}(f_{\mu }+J_{\mu }) \;+\;%
\frac{i}{2\eta }A_{\mu }\mathcal{F}^{\mu }+i\vartheta A_{\mu }J^{\mu
}\;\right) \;,  \label{sec-act1}
\end{equation}
\begin{equation}
J^{\mu }(x)=\int_{\gamma }dy^{\mu }\delta ^{(3)}(x-y)
\makebox[.5in]{,}\eta \vartheta^2=2\pi\;,  \label{j}
\end{equation}
the effective action turns out to be,
\begin{eqnarray}
\Gamma_{\gamma} &=&i\pi \int d^3x\, J_\mu  (\varepsilon\partial)^{-1}_{\mu \nu}\, J_\nu
+\frac{1}{2}(1-\vartheta \eta )^{2}\int d^{3}x\, J_{\mu }\frac{\hat{O}}{1-\eta ^{2}\hat{O}%
^{2}\partial ^{2}}\, J_{\mu }  \nonumber \\
&&+\frac{i}{2\eta }(1-\vartheta \eta )^{2}\int d^{3}x\,
J_{\mu }\frac{\hat{O}^{2}\partial^2}{1-\eta ^{2}\hat{O}^{2}\partial^{2}} (\varepsilon\partial)^{-1}_{\mu \nu}\, J_{\nu}\;,
\label{jj}
\end{eqnarray}
where $(\varepsilon\partial)^{-1}_{\mu \nu}$ is the Green function for the operator $(\varepsilon \partial)_{\mu \nu}$.

For large mass free fermions, the leading terms of the induced action are \cite{ox-so},
\begin{equation}
\Gamma _{\gamma }\sim \lambda mL+ i\pi S_{\gamma }
\makebox[.5in]{,}S_{\gamma}=\int d^3x\, J_\mu  (\varepsilon\partial)^{-1}_{\mu \nu} J_\nu\;,
\label{n-r}
\end{equation}
that is, Polyakov's Bose-Fermi transmutation occurs \cite{p,ha} and the vortex propagator turns out to be that of a
spin one half fermionic excitation,
\begin{equation}
\int d^{3}p\frac{1}{\sigma ^{\mu }p_{\mu }+\lambda m}e^{ip(x_{1}-x_{2})}\; ,
\label{dirac}
\end{equation}
where $\sigma ^{\mu }$ are the Pauli matrices; for more details, see ref. \cite{ox-so} and references therein.

\section{Analysis of the stability of vortex modes}
\label{analisis}

Now, when interactions are turned on, we will first search for possible
infrared divergences of the vortex energy, similar to those occurring in the single soliton sector in the $(1+1)D$ case.
The vortex energy can be read from eq. (\ref{jj}) as
the action per unit length for a straight string running along the euclidean time direction,
\begin{equation}
E=\frac{1}{2}(1-\vartheta \eta )^{2}\int \frac{d^{2}{\mathbf k}}{(2\pi )^{2}}\,\frac{%
\mathbf{\tilde{O}}}{1+\eta ^{2}\mathbf{\tilde{O}}^{2}\mathbf{k}^{2}}\;,%
\label{ener}
\end{equation}
where the quantities in boldface correspond to the two-dimensional
projection $k\rightarrow (0,\mathbf{k})$. Let us consider the class of interactions defined by
eq. (\ref{alfa}). It is easy to see that now, in the $(2+1)D$ case, there is no $\alpha$ that can render the integral in eq. (\ref{ener}) infrared divergent. This is clear if ${\cal P}$ is such that $\mathbf{\tilde{O}}$ tends to zero or a finite constant, when $|{\mathbf k}|\to 0$. In the case where $\mathbf{\tilde{O}}$ diverges faster than $|\mathbf{k}|^{-1}$, the infrared behavior of the integrand is proportional to
$(\mathbf{\tilde{O}}\mathbf{k}^{2})^{-1}$, which is infrared integrable in two dimensional momentum space. Finally, when $\mathbf{\tilde{O}}$ diverges like $|\mathbf{k}|^{-1}$ or slower, the integrand is proportional to $\mathbf{\tilde{O}}$, which is also integrable.
Then, in $(2+1)D$,  there is no energy instability of the vortices.
This is understandable as in $(2+1)D$ the quasiparticle picture is quite robust.

The next step is to study a possible destabilization in the localization properties of the theory.
Interactions could induce a breakdown of the large distance behavior in eq. (\ref{n-r}); in this case, the derivation of a propagator which at large distances behaves as the free fermion propagator is not guaranteed. If this happens, instead of a dressed propagator for spin one half quasiparticles, we could be led to a different phase where the fermion spectrum is destabilized.

For this aim, we will study the effect of interactions, defined by eq. (\ref{modelo}), on the induced string action in eq. (\ref{jj}),
with $\hat{O}$ given in eq. (\ref{valoro}). Note that for $1<\alpha\leq 2$ in eq. (\ref{alfa}), the associated potential decays faster than the Coulomb $1/R$ potential, in particular, $\alpha=2$ corresponds to a contact potential. For $\alpha<1$, the potential decay is slower than the Coulomb potential; $\alpha =0$ corresponds to a logarithmic potential, $\alpha=-1$ gives a linear confining potential, etc.

Equation (\ref{jj}) can also be written in terms of the associated kernels, for instance, the integral in the second term can be written
as,
\begin{equation}
\frac{1}{2} \int d^{3}x d^{3}y\, J_{\mu }(x) K\left( |x-y|^2\right) J_{\mu}(y)=\int d\tau d\tau'\, K\left( |x(\tau)-x(\tau')|^2\right)\;,
\label{kernel}
\end{equation}
\begin{eqnarray}
\lefteqn {K\left( |x-y|^2 \right))=\int d^3k\,\frac{\tilde{O}}{1+\eta ^{2}\tilde{O}^{2} k^{2}} e^{-ik(x-y)}}\nonumber \\
&&=\frac{4\pi m^2}{r}\int_0^\infty dv\;\frac{v A(v)}{1+\eta ^{2} v^2 A^2(v)} \sin(v r)\;,
\label{K}
\end{eqnarray}
where $x(\tau)$ is a parametrization of $\gamma$ and we have used that the theory is rotationally invariant as the nonlocal operators
depend on $\partial^2$. We have also introduced the dimensionless variable, $r=m|x-y|$, and defined,
\begin{equation}
A(v)= 1+\frac{\lambda m^{\alpha-1}}{v^{2-\alpha}}\;.
\label{A}
\end{equation}

Considering the parametrization $\tau\in [0,L]$, where $L$ is the length of the curve and,
\begin{equation}
e_\mu e_\mu=1 \makebox[.5in]{,}
e_\mu=\frac{dx_\mu}{d\tau}\;,
\label{e}
\end{equation}
and expanding the argument of the kernel in eq. (\ref{kernel}), we obtain,
\begin{equation}
|x(\tau)-x(\tau')|^2=(\tau-\tau')^2-\frac{2}{4!} \dot{e}_\mu \dot{e}_\mu|_{\tau'}(\tau-\tau')^4 +\dots\;,
\label{expansion}
\end{equation}
where we have used that eq. (\ref{e}) implies $\dot{e}_\mu e_\mu=0$, and similar relations.

\begin{figure}
\epsfxsize=10 cm
\epsfysize=8 cm
\epsfbox{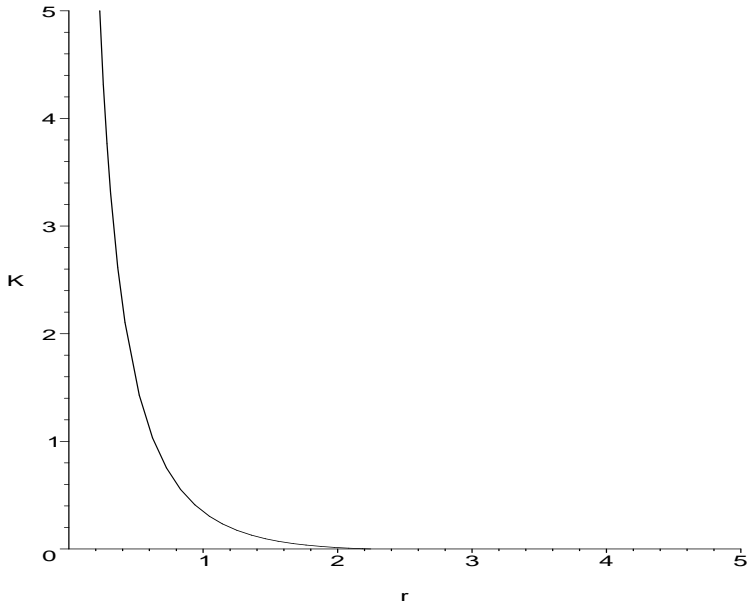}
\caption{vortex profile with $\alpha=-1$, $\lambda=2$}
\label{fig1}
\end{figure}

In general, when the interactions lead to a kernel in eq. (\ref{kernel}) containing a small localization scale $\delta$, in the sense that for $|\tau-\tau'|>\delta$ the kernel is suppressed, we can approximate the argument in eq. (\ref{kernel}) by the first term in eq. (\ref{expansion}), thus obtaining,
\begin{eqnarray}
\lefteqn{\int_0^L d\tau \int_0^L d\tau'\, K\left( |x(\tau)-x(\tau')|^2\right)\approx}\nonumber \\
&&\approx \int_0^L d\tau \int_0^L d\tau'\, K\left( (\tau-\tau')^2\right)\nonumber \\
&&\approx \int_0^L d\tau \int_{\tau-\delta}^{\tau+\delta}d\tau'\, K\left( (\tau-\tau')^2\right)\;,\nonumber \\
\end{eqnarray}
and changing variables $\tau'\to \tau' -\tau$,
\begin{eqnarray}
\lefteqn{\int_0^L d\tau \int_0^L d\tau'\, K\left( |x(\tau)-x(\tau')|^2\right)\approx}\nonumber \\
&&\approx \int_0^L d\tau \int_{-\delta}^{\delta}d\tau'\, K\left( {\tau'}^2\right)= {\rm const}\, L\;. \nonumber \\
\label{L}
\end{eqnarray}
In fig. (\ref{fig1}) we plot the kernel as a function of $r$ for $\alpha=-1$ and $\lambda=2$. This kernel turns out to be almost undistinguishable from the one obtained in the free fermion model, where $\hat{O}=1/m$ and the kernel is of Yukawa type with a localization length of the order of $\delta=1/m$. We have also verified that the kernel localization exists over the whole interval of $\alpha$'s, in spite of the fact that the associated potentials are in general long ranged.
Therefore, the term proportional to the string length in the large distance induced string action $S_\gamma$ in eq. (\ref{n-r}) is never destabilized.

Finally, let us look for a possible destabilization of the spin one half term.
Firstly, we note that for free fermions the third term in eq. (\ref{jj}) is irrelevant at large distances, when compared with the first spin one half term. However, if the interaction is such that $\tilde{O}^{2} k^{2}$ is nonzero at $k=0$, the coefficient for the spin one half action will
be changed at large distances and the free fermion propagator could be destabilized. According to eqs. (\ref{alfa}) and (\ref{valoro}), this happens when the associated potential is $R^{-\alpha}$, $\alpha \leq 1$, ranging from a Coulomb up to confining behaviors.

\section{Discussions \label{discusion}}
In this paper we have compared the effect of interactions on the asymptotic fermion spectrum in $(1+1)D$ and $(2+1)D$.
For this aim, we have used the identification of the fermionic excitations with finite energy topological configurations in the dual bosonized theory.

As is well known, in the case of $(1+1)D$ massive fermions, interactions associated with linear potentials destabilize the asymptotic fermionic states in the sense that the energy of a soliton configuration in the dual theory becomes infrared divergent, thus decoupling from the spectrum. This also occurs for any interaction leading to confining interparticle potentials.

On the other hand, we have shown that in $(2+1)D$ the vortex energy is always infrared finite. This is another way of understanding
that the free fermionic asymptotic spectrum in higher dimensions is quite robust.
However, we have seen that an interesting effect of interactions on the spin degrees of the model takes place. This comes about from the representation of the fermion propagator as a vortex correlator in the dual theory.
We have verified that for interactions associated with potentials $R^{-\alpha}$, $\alpha \leq 1$, there is a term in the induced action for the vortex worldline that competes with the spin one half Polyakov term at large distances. These potentials range from a Coulomb up to confining behaviors
($\alpha=-1$ corresponds to a linear interparticle potential). It would be interesting to investigate the lattice version of the fermionic theory (\ref{KF}) to see the cases where this competition is realized as a destabilization of the pole in the fermion propagator.

\section{Acknowledgments}

The Conselho Nacional de Desenvolvimento Cient\'{\i }fico e Tecnol\'{o}gico
CNPq-Brazil, the Funda{\c{c}}{\~{a}}o de Amparo {\`{a}} Pesquisa do Estado
do Rio de Janeiro (Faperj) and the SR2-UERJ are acknowledged for the
financial support.

\end{document}